\documentclass[twocolumn,aps,floats,floatfix,nofootinbib,prd,superscriptaddress,tightenlines]{revtex4}

\begin{document}

\title{Light scalar at LHC: the Higgs or the dilaton?}

\author{Walter D. Goldberger}
\affiliation{Department of Physics, Yale University, New Haven, CT 06520}

\author{Benjamin Grinstein}
\affiliation{Physics Department, University of California at San Diego, La Jolla, CA 92093}

\author{Witold Skiba}
\affiliation{Department of Physics, Yale University, New Haven, CT 06520}

\begin{abstract}
It is likely that the LHC will observe a color- and charge-neutral scalar whose decays are consistent with those of the Standard Model (SM) Higgs boson. The Higgs interpretation of such a discovery is not  the only possibility.   For example, electroweak symmetry breaking (EWSB) could be triggered by a spontaneously broken, nearly conformal sector. The spectrum of states at the electroweak scale would then contain a narrow scalar resonance, the pseudo-Goldstone boson of conformal symmetry breaking, with Higgs-like properties.   If the conformal sector is strongly coupled, this pseudo-dilaton may be the only new state accessible at high energy colliders.    We  discuss the prospects for distinguishing this mode from a minimal Higgs boson at the LHC and ILC.  The main discriminants between the two scenarios are (1) cubic self-interactions and (2) a potential enhancement of couplings to massless SM gauge bosons.   A particularly interesting situation arises when the scale $f$ of conformal symmetry breaking is approximately the electroweak scale $v\simeq 246~\mbox{GeV}$.    Although in this case the LHC may not be able to tell apart a pseudo-dilaton from the Higgs boson, the self-interactions differ in a way that depends only on the scaling dimension of certain operators in the conformal sector.  This opens the possibility of using dilaton pair production at future colliders as a probe of EWSB induced by nearly conformal new physics.

\end{abstract}

\maketitle

In the absence of an explicit sector that breaks electroweak gauge invariance, the interactions of SM gauge bosons and fermions are approximately conformal down to the QCD scale.  Thus the question of what triggers gauge symmetry breaking in the SM is inevitably tied to the dynamical breaking of scale invariance.  This is why the Higgs mode and the dilaton mode~\cite{GildenerWeinberg}, the pseudo-Goldstone boson of spontaneously broken scale invariance, can have such similar properties.

For example, in the minimal SM with a single Higgs doublet, the breaking of $SU(2)_L\times U(1)_Y$ gauge invariance is triggered by \emph{explicit} scale symmetry breaking, through the addition of the (fine tuned) Higgs mass operator.   In this case, the scale $f$ of conformal breaking is exactly equal to the scale $v\simeq 246~\mbox{GeV}$ of the electroweak interactions.   In addition, if the explicit scale breaking term is small in units of $f=v$, the resulting light Higgs can be identified with the dilaton $\chi(x)=\sqrt{H^\dagger H}(x)$.  The couplings of this light mode to the rest of the SM satisfy the usual soft Higgs theorems~\cite{softHiggs}, which in this language are a consequence of the Ward identities of approximate scale symmetry.

If the mechanism of EWSB is more subtle than a simple Higgs scalar, the relation between gauge and scale symmetry breaking is likely to be less direct.   For example, if the EWSB sector is strongly coupled,    the conformal breaking scale $f$ need not be the same as $v$, nor do the states responsible for EWSB (and for unitarizing gauge boson scattering) have to be identical to the states that arise from the conformal sector of the theory.      

It is not difficult to find explicit examples where this is the case.   In fact, a wide range of well motivated extension of the SM contain an EWSB sector that is strongly interacting and nearly conformal.   The general picture is that in these theories, spontaneous breaking of scale invariance at a scale $\Lambda_{CFT}\sim 4\pi f$ triggers EWSB at an energy scale $\Lambda_{EW} \sim  4\pi v\leq \Lambda_{CFT}$.   Models of this type can be realized either as 4D strongly coupled gauge theories, as is the case in the original theories of walking technicolor~\cite{walking}, or via AdS/CFT as Randall-Sundrum 5D warped geometries~\cite{RS} with EWSB through either Higgs VEVs~\cite{RShiggs} or boundary conditions~\cite{RSbound}.   See also~\cite{LutyOkui}.

In such scenarios, the EWSB sector may not correspond to new, light resonances that can be directly seen at the LHC.   Rather, it is realized non-linearly at low energies, and the states responsible for unitarizing gauge boson scattering amplitudes are heavy compared to the weak scale.   For example if the underlying theory is strongly coupled, new states with masses of order $\Lambda_{EW}\sim 4\pi v$ are generic. Such heavy states are typically broad and therefore difficult to detect experimentally.

Despite the lack of light states directly tied to the mechanism of EWSB, the theory may still posses a light electroweak singlet dilaton field $\chi(x)$.   In the limit where conformal symmetry is recovered, the dilaton is exactly massless.   Therefore its mass is naturally small\footnote{There are arguments in the literature, for example in Ref.~\cite{Bardeen:1985sm,nolightdilaton}, against a light dilaton in gauge theories.   Such studies obtain properties of the spectrum by analyzing solutions to the Schwinger-Dyson equations of the theory and likely do not accurately capture properties of all gauge theories.   For example, large $N$ gauge theories obtained through the AdS/CFT correspondence can have a light dilaton in the spectrum.}, proportional to the scale $f$ times the parameter that controls deviations from exact scale invariance, and the pseudo-dilaton may well be the first state associated with new physics at the LHC. 

Because a light Higgs doublet can also be described as a pseudo-dilaton, differentiating the new physics from the minimal SM at colliders may be difficult.   Indeed, exactly the same low energy theorems that determine the interactions of the Higgs with the rest of the SM apply as well to the couplings of a pseudo-dilaton that arises from an underlying strong sector.    By non-linearly realized scale invariance, the tree level pseudo-dilaton couplings are obtained from those of the Higgs boson by replacing the electroweak scale $v$ with the scale $f$ of conformal symmetry breaking.  In general the scale $f$ may differ from $v$, so that the unitarization of Standard Model amplitudes is only partial.   At the loop level, the dilaton also has couplings to massless SM gauge bosons, but their strength is model dependent.   However, if $f$ is close to $v$ and no other light states appear, it would be nearly impossible to distinguish the new strong dynamics from the minimal Higgs model.   Finally,  the self-couplings depend on the dimension of operators that explicitly break the conformal symmetry, and this provides an opportunity to differentiate the dilaton from the Higgs.

In the next section we describe our setup,  which consists of the SM with nonlinearly realized electroweak symmetry coupled to a light dilaton arising from a conformal theory.    There, we discuss the couplings of the dilaton to SM fields.   Assuming that the SM is embedded in the conformal sector, we obtain the dilaton couplings to massless SM gauge bosons,  finding a large enhancement relative to the analogous SM Higgs couplings. Unfortunately, this does not constitute a tell tale sign of the dilaton since the perturbative SM gauge interactions can be non-conformal at high scales without spoiling the dynamics of a strongly interacting conformal sector.   In addition, Higgs couplings to photons and gluons can be radiatively induced by new heavy states, making an unambiguous comparison of the  Higgs and dilaton couplings difficult.  We also show that, in certain limits, the dilaton cubic self coupling can be calculated in terms of its mass and the scale of conformal symmetry breaking. This coupling, although difficult to measure, is perhaps the most interesting probe of the conformal sector. 

In section~\ref{sec:colliders} we briefly discuss collider phenomenology. We first address current bounds inferred by LEP experiments, then turn to LHC prospects, and finally comment on the measurement of the dilaton self coupling that could be achieved at a linear collider.

\section{Setup}

We assume that the light degrees of freedom consist of the SM gauge bosons and fermions as well as an electroweak neutral scalar, the dilaton, whose mass is protected by approximate scale invariance.   All other states responsible either for conformal or electroweak breaking are taken to be roughly heavier than a scale $\Lambda_{EW}\sim 4\pi v$.       The interactions among the light fields are described by a low-energy effective Lagrangian with non-linearly realized $SU(2)_L\times U(1)_Y$ and (approximate) conformal invariance.

\subsection{Conformal sector}

In this section we briefly recall the standard lore on broken scale invariance.   Conformal invariance in a field theory can be broken by both classical (couplings with non-zero mass dimension) and quantum (dependence on a renormalization scale $\mu$) effects.   The manner in which scale invariance is broken can be read off the RG equations of the theory.

A heuristic way of understanding the effects of scale transformations on the theory can be obtained by writing the Lagrangian in a basis of anomalous dimension eigen-operators,
\begin{equation}
\label{eq:L}
{\cal L} = \sum_i g_i (\mu) {\cal O}_i(x),
\end{equation}
with $[{\cal O}_i]=d_i$.   The content of the RG equations is summarized by assigning  the transformations laws
\begin{eqnarray}
\nonumber
{\cal O}_i(x) &\rightarrow& e^{\lambda d_i} {\cal O}_i( e^{\lambda} x),\\
\nonumber
\mu &\rightarrow & e^{-\lambda} \mu,
\end{eqnarray}
under scale transformations $x^\mu\rightarrow e^{\lambda} x^\mu$.   This gives
\begin{equation}
\label{eq:dL}
\delta {\cal L} = \sum_i  g_i(\mu) (d_i + x^\mu\partial_\mu) {\cal O}_i(x) + \sum_i {\beta_i}(g) {\partial\over \partial g_i} {\cal L},
\end{equation}
where $\beta_i(g)=\mu  \partial g_i(\mu)/\partial\mu$.   From this one obtains the divergence of the scale current $S^\mu = {T^\mu}_\nu x^\nu$,
\begin{equation}
\label{eq:div}
\partial_\mu S^\mu = {T^\mu}_\mu =  \sum_i  g_i(\mu) (d_i-4) {\cal O}_i(x) + \sum_i {\beta_i}(g) {\partial\over \partial g_i} {\cal L},
\end{equation}
This implies in particular the standard result that if $d_i=4$ and $\beta_i=0$, the theory is scale invariant.

Given Eq.~(\ref{eq:L}), a simple way of incorporating non-linearly realized scale invariance is to add a field $\chi(x)$ that serves as a conformal compensator.     Assigning the scale transformation law
$$
\chi(x) \rightarrow e^{\lambda} \chi(e^{\lambda} x),
$$
we simply need to make the replacement
$$
g_i(\mu)\rightarrow g_i\left(\mu {\chi\over f}\right) \left({\chi\over f}\right)^{4-d_i},
$$
in Eq.~(\ref{eq:L}).   Here $f=\langle\chi\rangle$ is the order parameter for scale symmetry breaking, determined by the dynamics of the underlying strong sector.   The Goldstone boson associated with conformal symmetry breaking is parameterized as
\begin{equation}
\chi(x) = fe^{\sigma(x)/f},
\end{equation} 
which transforms non-linearly under scale transformations, $\lambda:  \sigma(x)/f\rightarrow \sigma(e^{\lambda} x)/f+\lambda$.      However, a more convenient parameterization for the fluctuations about the VEV is ${\bar \chi}(x) =  \chi(x)-f.$  Expanding about $\langle\chi\rangle=f$, one gets the standard result
\begin{equation}
{\cal L}_\chi = {1\over 2}\partial_\mu{\bar\chi} \partial^\mu {\bar\chi}  + {\bar\chi\over f} {T^\mu}_\mu+\cdots,
\end{equation}
with ${T^{\mu}}_\mu$ as in Eq.~(\ref{eq:div}).

\subsection{Electroweak sector}

A convenient, model-independent description of a strongly interacting Higgs sector is in terms of the electroweak chiral Lagrangian~\cite{ewchiral}.   Introducing a  $2\times 2$ unimodular matrix field $U(x)$, the dynamics of the EWSB sector at energies below  $\Lambda_{EW}\sim 4 \pi v\simeq 1~\mbox{TeV}$ is given by
\begin{equation}
\label{eq:EW}
{\cal L}_{EW} = {\cal L}_{\chi EW} + {\cal L}_\psi + {\cal L}_Y,
\end{equation}
with 
\begin{equation}
\label{eq:chiEW}
{\cal L}_{\chi EW} = -{1\over 4} (B_{\mu\nu})^2 -{1\over 2}\mbox{tr}W_{\mu\nu}^2 + {1\over 4} v^2 \mbox{tr} D_\mu U^\dagger D^\mu U+ \cdots,
\end{equation}
where the covariant derivative of $U(x)$ is
\begin{equation}
D_\mu U = \partial_\mu U +  i g_1 B_\mu U{\tau_3\over 2} - i g_2 {\vec W}_\mu \cdot {{\vec \tau}\over2}U,
\end{equation}
and
\begin{equation}
{\cal L}_Y = - {\bar Q}_L U m_q q_R - {\bar L}_L U m_{\ell} {\ell}_R +\mbox{h.c.}
\end{equation}
where $m_q/v,$ $m_{\ell}/v$ are quark and lepton Yukawa matrices\footnote{We have written the right-handed fermions as custodial $SU(2)$ doublets, so that $m_{q,\ell}$ is a $2\times 2$ diagonal matrix of $3\times 3$ blocks, with the lower block of $m_{\ell}$ set to zero.}.   The term ${\cal L}_\psi$ contains the usual fermion kinetic energy operators.

In the unitary gauge, $U=1$, ${\cal L}_{\chi EW}$ above describes the kinetic and mass terms for the $SU(2)_L\times U(1)_Y$ gauge fields.   Terms omitted in Eq.~(\ref{eq:chiEW}) are higher derivative operators that encode the various precision electroweak parameters with coefficients that scale as inverse powers of the scale $\Lambda_{EW}$.   We simply assume that these coefficients are adjusted to be consistent with the measured experimental values of the electroweak observables.   We have also neglected an additional custodial $SU(2)$ violating two-derivative operator whose coefficient is experimentally known to be small.

It is clear that the gauge boson and fermion mass terms include the coupling of gauge fields to the dilaton as, the replacement $v\rightarrow v\chi/f$ makes Eq.~(\ref{eq:EW}) formally scale invariant.   Expanding about $\langle \chi\rangle =f$ gives the couplings of the dilaton to the SM gauge bosons and fermions at tree level
\begin{eqnarray}
\label{eq:SMcouplings}
\nonumber
{\cal L}_{\chi,SM} &=&  \left({2\bar\chi\over f}+{\bar\chi^2\over f^2}\right) \left[ m_W^2 W^+_\mu {W^-}^\mu +{1\over 2}m_Z^2 Z_\mu Z^\mu\right] \\
& & {}  + {\bar\chi\over f}\sum_\psi m_\psi {\bar \psi} \psi,
\end{eqnarray}
which are identical in form to the couplings of a minimal Higgs boson.

\subsection{Dilaton self couplings}

In the limit of exact scale invariance $\chi$ is derivatively self-coupled.   Ignoring for the time being terms that explicitly break the symmetry, self-interactions of the dilaton take the form
\begin{equation}
{\cal L}_\chi =  {1\over 2}\partial_\mu\chi\partial^\mu\chi + {c_4\over (4\pi\chi)^4} \left(\partial_\mu\chi\partial^\mu\chi\right)^2+\cdots,
\end{equation}
where the constant $c_4\sim{\cal O}(1)$ depends on the details of the underlying CFT.   The inverse powers of $\chi$ are necessary to ensure that ${\cal L}_\chi$ transforms correctly under scalings.

In addition, the theory may possess explicit sources of scale symmetry breaking.   For example, suppose that conformal invariance is broken by the addition of an operator ${\cal O}(x)$ with scaling dimension $\Delta_{\cal O}\neq 4$  to the Lagrangian,
\begin{equation}
{\cal L}_{CFT}\rightarrow {\cal L}_{CFT} + \lambda_{\cal O} {\cal O}(x).
\end{equation}
It is straightforward to include this pattern of symmetry breaking by the introduction of a spurion field into the low-energy effective theory.    This spurion constrains the non-derivative interactions of $\chi(x)$ to be of the form~\cite{RattazziZaffaroni}
\begin{equation}
\label{eq:spur}
V(\chi) = \chi^4 \sum^\infty_{n=0} c_n(\Delta_{\cal O}) \left({\chi\over f}\right)^{n(\Delta_{\cal O}-4)},
\end{equation}   
where the coefficients $c_n\sim\lambda^n_{\cal O}$ depend on the dynamics of the underlying CFT.  By assumption, this dynamics must be such that $V(\chi)$ is minimized at $\langle\chi\rangle=f$ with $m^2_\chi=d^2 V(\langle\chi\rangle)/d\chi^2 >0$. In general,  the coefficients $c_n$ are functions of the scaling dimension, which we assume are non-singular in the limit $\Delta_{\cal O}\rightarrow 4$.     

It is not possible to make detailed predictions without knowledge of the coefficients $c_n$ in $V(\chi)$
unless there exists a small expansion parameter.  Here we are interested in the case where the explicit conformal breaking term above is small.   This can be either because the operator ${\cal O}$  is nearly marginal ($|\Delta_{\cal O}-4|\ll 1$), as is the case in walking technicolor theories or RS models stabilized by the scenario of~\cite{GW}, or because the coefficient $\lambda_{\cal O}$ is chosen to be small in units of $f$, as in the case of the minimal Higgs model.   If this happens, it is possible to obtain definite expressions for the dilaton self-couplings once the parameters $m$ and $f$ are fixed.   We find that the potential is
\begin{equation}
V({\bar \chi}) = {1\over 2} m^2 {\bar \chi}^2 + {\lambda\over 3!}  {m^2\over f}{\bar \chi}^3+\cdots,
\end{equation}
where $m^2\ll f^2$ is proportional to the small symmetry breaking parameter:   $m^2/f^2\propto \lambda_{\cal O}$ for $\lambda_{\cal O}\ll 1$ (in units of $f$) and $\Delta_{\cal O}$ arbitrary, or  $m^2/f^2\sim |\Delta_{\cal O}-4|$ for $|\Delta_{\cal O} -4| \ll 1$ and  $\lambda_{\cal O}$ of arbitrary size.   The cubic coupling is given by
\begin{equation}
\label{eq:cubic}
\lambda = \left\{\begin{array}{ll} (\Delta_{\cal O}+1)+{\cal O}(\lambda_{\cal O}) & {\rm when}\  \lambda_{\cal O}\ll 1,\\
\\
  5+ {\cal O}\left( |\Delta_{\cal O}-4| \right) & {\rm when}\  |\Delta_{\cal O}-4|\ll 1,
\end{array}\right.
\end{equation}
and is in principle a probe of the scaling dimension of the operator responsible for scale symmetry breaking.   We do not expect scale symmetry breaking to occur if ${\cal O}(x)$ is an IR irrelevant perturbation.    This implies the bound $\lambda\leq 5$ that is saturated near marginality.  In addition, the conformal algebra together with unitarity implies $\lambda\geq 2$.  Moreover,  for $\Delta_{\cal O}=2$ and $\lambda_{\cal O}\ll f^2$  the result in Eq.~(\ref{eq:cubic}) reproduces the usual Higgs trilinear coupling. Note that when $|\Delta_{\cal O}-4|\ll 1$ the entire potential for $\chi$, up to corrections of order  $(\Delta_{\cal O}-4)^2$, is calculable. In fact, it is
\begin{displaymath}
 V(\chi)=  \frac{1}{16} \frac{m^2}{f^2} \chi^4 \left[4 \ln\frac{\chi}{f} -1 \right] +  {\cal O}\left( |\Delta_{\cal O}-4|^2 \right).
\end{displaymath}

Finally, one may worry that radiative corrections could spoil our predictions for the dilaton self couplings.    For example, given the couplings in Eq.~(\ref{eq:SMcouplings}) one might naively expect that top quark loops could generate radiative corrections to $m^2$ of magnitude 
$$
\delta m^2\sim {m^2_t \Lambda^2 \over 16\pi^2 f^2},
$$
where $\Lambda$ is an ultraviolet cutoff.   Likewise, the cubic coupling receives a one-loop linearly divergent contribution of the form $m^3_t \Lambda/(16\pi^2 f^3)$.   In fact, such corrections are absent, simply by the assumption that the theory is invariant under scale transformations in the limit $\lambda_{\cal O}\rightarrow 0$.     Physically this means that for any SM corrections to $V(\chi)$, there must be corresponding cancellations from states at high scales to ensure scale invariance.    A way to implement this in the low energy effective theory is to formally make the UV cutoff have explicit dependence on the field $\chi,$
$$\Lambda\rightarrow \Lambda {\chi\over f}.
$$
This is consistent with the interpretation of $\chi$ as a conformal compensator.   Making this replacement ensures that the only corrections to $V(\chi)$ are quartic in the field.   Since an arbitrary quartic term has been included in Eq.~(\ref{eq:spur}), this means that our prediction for the cubic self-coupling of ${\bar\chi}$ is stable against loop corrections.

\subsection{Couplings to massless gauge bosons}
\label{sec:Higgs-massless}

In the SM, Higgs couplings to the top quark and the massive gauge bosons induce the couplings $H\gamma \gamma$ and $Hgg$.  The same mechanism induces the dilaton couplings $\chi \gamma \gamma$ and $\chi gg$.   Because these processes are generated by loop effects, these couplings are also sensitive to contributions from heavy particles present in any extension of the SM.   Since these couplings are crucial for collider phenomenology we derive them here and show that the dilaton coupling to gluons and photons can be significantly enhanced under very mild assumptions about high scale physics.   

Let us begin by recalling the situation for the SM loop induced Higgs couplings to gluons.    The logic is identical for the couplings to two photons.    One way of obtaining the $Hgg$ vertex is to compute the effective action for background color and Higgs fields.    If we are interested in the coupling induced by heavy particles, we can take the Higgs background to be spacetime independent.    Then the relevant term is of the form
\begin{equation}
\Gamma[A,H]= {1\over 4}\int_q {\hat G}_{\mu\nu}^a(-q) \Pi(q^2,H)  {\hat G}^{\mu\nu a}(q) +\cdots,
\end{equation}
where the vacuum polarization function $\Pi(q^2,H)$ includes all heavy particle loops.   Taking for illustration the one-loop contribution of fermions only (although all particles charged under $SU(3)_c$ must be included to make the answer independent of renormalization scale $\mu$), $\Pi(q^2,H)$ is given by 
\begin{eqnarray}
\label{eq:effaction}
\nonumber
\Pi(q^2,H) =  \frac{1}{g^2(\mu)} - \frac{4}{(4 \pi)^2} \sum_i   \int^1_0 dx \, x (1-x)\\
{} \times    \ln\left[\frac{x (1-x) q^2 +{2 m^2_i} H^\dagger H/v^2}{\mu^2} \right].
\end{eqnarray}
For Higgs processes, $q^2$ is taken to be typically of order $m^2_h$.    In general, the sum runs over all particles, however when $m_i^2  \ll q^2 \approx m_h^2$ the logarithm is dominated by the $q^2$ term and is independent of the Higgs field, so as expected light particles induce negligible $Hgg$ couplings. For heavy particles, defined as $m_i  \gg m_h$, we can neglect the $q^2$ term, and the Higgs couples to gluons through an operator
\begin{equation}
\label{eq:hgg}
{\cal L}_{hGG} =  \frac{\alpha_s}{8 \pi}  \sum_{i} b^i_0 \frac{h}{v} ({G}_{\mu\nu}^a)^2,
\end{equation}
where we have expanded about the Higgs VEV and $G^a_{\mu\nu}$ is the canonically normalized gluon field strength.   The sum runs over the heavy fields only and $b^i_0$ is the contribution of each heavy particle to the one-loop QCD beta function, normalized as $\beta_i(g)=b^i_0 g^3/16 \pi^2$.  As expected, this result is independent of the heavy masses. For example, for a heavy fermion like the top quark $b_0=2/3$.   

This well known result is modified by the loop contributions of  other heavy particles which are not part of the SM.    In addition to inducing the term in Eq.~(\ref{eq:hgg}), such heavy states can also generate new dimension-six operators of the form, e.g., 
\begin{equation}
{\cal L}_{hGG}  \supset {\alpha_s\over 4\pi} c_{hg} H^\dagger H ({G}_{\mu\nu}^a)^2,
\end{equation}
which depending on the size of the coefficient $c_{hg}$ can significantly modify the properties of a light Higgs boson~\cite{ManoharWise}.

The dilaton couplings to massless gauge bosons can be simply obtained by making the replacement 
\begin{displaymath} 
\frac {2 m^2_i}{v^2} H^\dagger H  \longrightarrow 
    \frac {m^2_i}{f^2} \chi^2,
\end{displaymath}
in Eq.~(\ref{eq:effaction}).   Again, one can split the sum over all colored particles into sums over light and heavy states, where the dividing scale is given by the dilaton mass.   Note that if one assumes that QCD is fully embedded in the conformal sector, one can make UV insensitive predictions, since by conformal invariance    
\begin{displaymath}
   \sum_{\rm light} b_0+ \sum_{\rm heavy} b_0=0.
\end{displaymath}
Thus the effective coupling is 
\begin{equation}
\label{eq:chigg}
 {\cal L}_{\chi gg} =  - \frac{\alpha_s}{8 \pi}  b_0^{\rm light} \frac{\bar{\chi}}{f} ({G}_{\mu\nu}^a)^2,
\end{equation}
where $b_0^{\rm light}=-11 + \frac{2}{3} n_{\rm light}$.   The number of light fermions, $n_{\rm light}$, is either $n_{\rm light}=5$ if the dilaton is lighter than the top quark, or $n_{\rm light}=6$ otherwise.  Eq.~(\ref{eq:chigg}) has a non-perturbative generalization
\begin{equation}
{\cal L}_{\chi gg} =  - {\beta(g)\over 2 g} \frac{\bar{\chi}}{f} ({G}_{\mu\nu}^a)^2,
\end{equation}   
where $\beta(g)$ is the beta function including particles lighter than the dilaton mass.   For collider applications Eq.~(\ref{eq:chigg}) is sufficient, however.   It indicates about a tenfold increase of the coupling strength compared to that of the SM Higgs, which could have profound consequences at the LHC.   Unlike the Higgs case, corrections to this result from higher dimension operators are negligible.    For example, one might consider operators such as
\begin{equation}
{\cal L}_{\chi gg} \supset g^2_s{c_{\chi g}\over (4\pi\chi)^2} D_\alpha G^a_{\mu\nu}D^\alpha G^{\mu\nu a}.
\end{equation} 
However, such operators are suppressed by powers of $m^2/f^2\ll 1$ relative to the terms coming from the conformal anomaly.

\section{Collider physics}
\label{sec:colliders}

The couplings of the dilaton at energies below the scale $4\pi f$ are given by 
\begin{eqnarray}
\nonumber
{\cal L}_{\chi} &=& {1\over 2} \partial_\mu {\bar \chi} \partial^\mu {\bar\chi} -{1\over 2} m^2 {\bar\chi}^2 + {\lambda\over 3!} {m^2\over f} {\bar\chi}^3 + {\bar\chi\over f}\sum_\psi m_\psi {\bar \psi} \psi\\
\nonumber
& & {} + \left({2\bar\chi\over f}+{\bar\chi^2\over f^2}\right) \left[ m_W^2 W^+_\mu {W^-}^\mu +{1\over 2}m_Z^2 Z_\mu Z^\mu\right]\\
& & {} + {\alpha_{EM} \over 8 \pi f} c_{EM}{\bar\chi} (F_{\mu\nu})^2+ {\alpha_s  \over 8\pi f} c_G {\bar\chi} (G^a_{\mu\nu})^2,
\end{eqnarray}
where the coefficients $c_{EM},$ $c_{G}$ were discussed in the previous section.    For example, if electromagnetic and strong interactions are embedded in the conformal sector at high scales,
\begin{equation}
\label{eq:cubic}
c_{EM} = \left\{\begin{array}{ll} -17/9 & {\rm when} \, m_W<m<m_t,\\
\\
-11/3 & {\rm when}\,  m>m_t,
\end{array}\right.
\end{equation}
 while $c_{G}=11-2 n_{\rm light}/3$, where $n_{\rm light}$ is the number of quarks lighter than the dilaton.

Given the similarity to minimal Higgs physics, it is possible to use existing studies of Higgs properties at colliders to understand the physics of a light dilaton as a function of the model parameters $m$, $f$, and the couplings $\lambda$, $c_{EM}$, $c_G$.

\subsection{LEP bounds}

At LEP, the main production channel for dilaton production is, as for the Higgs, associated production with a  virtual $Z$ boson, $e^+e^-\rightarrow HZ^*$.  The cross section for dilaton production
is suppressed by a factor $(v/f)^2$ relative to the corresponding Higgs cross section at the same mass.
The LEP collaborations have combined their data to search for the Higgs, including a search for Higgs particles with an anomalous (non-SM) $HZZ$ coupling~\cite{Higgsworkinggroup}.   This result is immediately applicable to the bounds on the dilaton mass and coupling.   

Figure 10 in Ref.~\cite{Higgsworkinggroup} summarizes the bound on the dilaton mass and decay
constant, where in our case $\xi^2=(v/f)^2$.   Roughly, the dilaton with mass $90\ {\rm GeV}<m <110\ {\rm GeV}$ is excluded if $(v/f)^2>0.1$ and with mass $12\ {\rm GeV}<m<90\ {\rm GeV}$
it is excluded for   $(v/f)^2>0.01$. These limits predominantly come from the $b\bar{b}$ decay
channel, which is kinematically suppressed below $12\ {\rm GeV}$. Other available decay channels have been employed for very light masses~\cite{OPAL}.   Values $m <12\ {\rm GeV}$ are
excluded if $(v/f)^2>0.1$; see Figure 5 in Ref.~\cite{OPAL}. 

The dilaton decay width into quarks and leptons is also suppressed by the factor $(v/f)^2$.   However, this discrepancy is not relevant for the LEP search as the branching ratios to fermions remain unchanged.  For $(v/f)^2<10^{-2}$, LEP is not able to detect the dilaton irrespective of its mass, while for $(v/f)^2>10^{-2}$ the suppression of the width is not observable.  In this latter case the dilaton decays very promptly and does not have displaced decay vertex. Therefore its signatures are identical to Higgs signatures.

\subsection{LHC}

There are four important production channels for the dilaton at hadron colliders: gluon fusion $gg\rightarrow \chi$, associated production with vector bosons $q\bar{q} \rightarrow W/Z + \chi$, vector boson fusion $qq \rightarrow qq + \chi$, and associated production with the top quark $gg,\ q\bar{q}\rightarrow t\bar{t} +\chi$.  The first process, $gg\rightarrow \chi$, is likely to be sensitive to new heavy states as we discussed in Sec.~\ref{sec:Higgs-massless}.  For example, assuming that QCD is embedded in the conformal sector, we find a large enhancement of the $\chi gg$ coupling.  This could easily overcome the suppression factor $v/f$ when compared with the $gg\rightarrow h$ cross section. The cross sections for the remaining process scale as $(v/f)^2$ compared to the Higgs production cross sections. Higher order QCD corrections, which are often sizable, do not alter the scaling of the cross sections with $(v/f)^2$  since each of these process contains just one vertex involving the dilaton. As we already discussed, the dominant branching ratios of the dilaton are the same as for the Higgs boson so most of the Higgs search strategies can be applied directly. The only caveat is in the mode $\chi\rightarrow \gamma \gamma$. The width of this decay is likely to be modified by physics beyond SM and scaling the results obtained for the Higgs may not be reliable. 

Given the simple scaling of the cross section we can estimate the reach of LHC as a function of the dilaton mass and the decay constant $f$. The statistical significance of the Higgs signal at ATLAS has been presented, for example, in Refs.~\cite{Djouadi,ATLASreach}. The significance of the dilaton signal can be obtained from the significance of the Higgs signal by rescaling  
\begin{displaymath}
    \left(\frac{S}{\sqrt{B}}\right)_\chi=  c_G^2 \frac{v^2}{f^2}  \left(\frac{S}{\sqrt{B}}\right)_{\rm Higgs},
\end{displaymath}
where we assume that the production cross section is dominated by the gluon fusion process.
It is easiest to discover a heavy dilaton when the decays to $WW$ and $ZZ$ dominate. Very crudely, with a $100~{\rm fb}^{-1}$ of integrated luminosity a discovery is possible when $c^2_G (v/f)^2 > 1/8$ for $m>160~{\rm GeV}$.  For a lighter dilaton the statistical significance decreases with mass, so if $c^2_G(v/f)^2\sim 1/10$ one may have to wait to collect about $300~{\rm fb}^{-1}$ worth of data for detection. For details see Figure 3.49 in Ref.~\cite{Djouadi}. 

In addition to discovery, the LHC will be able to measure the dilaton couplings to gauge bosons and the top quark by measuring event rates in different channels. Depending on the mass and the production channel one expects a 10\% to 30\% accuracy for the extraction of the couplings. The measurement of the cubic self coupling seems hopeless at the LHC even if $f\approx v$ and it only gets harder for $f>v$~\cite{Djouadi}. However, a large luminosity mode of the LHC or a VLHC could probe the cubic coupling~\cite{cubichadron}.
  
\subsection{ILC}
A linear collider with $\sqrt{s}=500-1000\ {\rm GeV}$ would provide an ideal environment for the
study of the dilaton and for distinguishing the dilaton from the Higgs.  A number of precision measurements can be performed.  See Ref.~\cite{Djouadi} for a comprehensive review.

First, the couplings to the gauge bosons and several branching ratios can be measured at a one to few percent level. In addition to determining $f$ it would be a test of whether or not different coupling are scaled by the universal factor $v/f$. 

Second, for $m >200\ {\rm GeV}$ and $f\approx v$, the dilaton would be broad enough for its total width to be observed directly. This would provide yet another check of the universal rescaling of the couplings relative to the Higgs.   For $f>v$ such a measurement is possible if $m> (f/v)^{2/3}\, 200\ {\rm GeV}$ since, in this mass range, the total width increases as mass cubed. (The Higgs mass reach of a linear collider is approximately $0.8 \sqrt{s}$.)

It would be fascinating if $f \approx v$ to a degree of accuracy that previously described coupling measurements would not distinguish the dilaton from the Higgs. The cubic coupling may then provide the only probe of how conformal symmetry is broken. If the conformal symmetry is broken by a nearly marginal operator we expect a slight enhancement of the cubic coupling by a factor of $5/3$. This is large enough to be probed by the ILC if the dilaton is light enough. The limiting factor for this measurement is the small production cross section of Higgs/dilaton pairs, which drops very rapidly with mass~\cite{cubicee}. For $f\approx v$, we can adopt the results of studies on Higgs pair production, which estimate that the cubic coupling can be measured to within, e.g., $20\%$ if $m=120 \ {\rm GeV}$ and to within $30\%$ if $m=140 \ {\rm GeV}$~\cite{Djouadi}.    Note that for $f\gg v$ the error in the cubic coupling determination should be multiplied by an additional factor of $(f/v)^2$ if the limitation to the measurement is assumed to be purely statistical and by $(f/v)^4$ under the assumption that it is background dominated.      Although neither of these extremes is likely to capture the true experimental situation, it does indicate that for $f$ significantly larger than $v$ the pair production signal is not an effective probe of dilaton physics.

\section{Concluding remarks}

We have considered a framework for physics at the TeV scale in which the breaking of electroweak symmetry is triggered by the dynamics of a strongly coupled, nearly conformal sector.   In scenarios of this sort (which include several well motivated proposals for physics beyond the SM), there may be a light scalar field, the pseudo-dilaton, which couples to ${{T_{SM}}^{\mu}}_{\mu}$ with strength inversely proportional to a symmetry breaking scale $f$.  It may in fact be the lightest new state seen at colliders.

Because a light Higgs boson can be also be thought of as an approximate dilaton, distinguishing the minimal SM from strong conformal dynamics may be difficult for $f$ close to $v$.   In this paper we have discussed the possible ways in which these two types of scenarios may be disentangled at colliders.   Most experimental signatures depend on just three parameters:  the dilaton mass $m$, the scale $f$, and for the dilaton self-couplings, the scaling dimension $\Delta$ of the operator that explicit breaks the conformal symmetry.   We find in particular that if $f$ is within roughly 10\% of the electroweak scale, LHC data is not sufficient to distinguish the dilaton from a minimal Higgs sector, and experimental determination of the self-coupling at the proposed ILC is required as well.     

Because the self-coupling is a direct probe of the mechanism of conformal symmetry breaking, it is important to determine the accuracy to which it can be measured at a linear collider.   This depends both on the dilaton mass and on the ratio $v/f$.    Our analysis of this has been at the order of magnitude level, adopting existing work on the Higgs cubic coupling.   It is important to do a more careful study that goes beyond the simple scaling arguments presented here.     It would also be interesting to work out the  more model dependent aspects of dilaton physics that have been neglected here.   For example, the dilaton couplings to the SM could be modified by mixing with the conformal sector, in which case the branching ratios for dilaton decays may deviate from those of the minimal Higgs.   Finally, our prediction for the dilaton self-couplings relies on the assumption that the conformal sector is explicitly broken by the addition of a single operator of dimension $\Delta\leq 4$, leading to a bound on the cubic coupling $\lambda\leq 5$.    It is possible that this bound can be relaxed with a more elaborate scale breaking sector involving several nearly-marginal operators, in which case dilaton pair production at a linear collider could be greatly enhanced.    The implications of this possibility need to be worked out in more detail.

\section{Acknowledgments}
We thank T. Appelquist for discussions.  This work has been supported in part by grants from the US Department of Energy. WG is supported by the Outstanding Junior Investigator award and grant DE-FG-02-92ER40704, BG by grant DE-FG03-97ER40546, while WS by  grant DE-FG-02-92ER40704.

\end{document}